%% file: FastCat__MAIN.tex
\documentclass[acmlarge]{acmart}

\renewcommand\footnotetextcopyrightpermission[1]{} 

\AtBeginDocument{%
  \providecommand\BibTeX{{%
    \normalfont B\kern-0.5em{\scshape i\kern-0.25em b}\kern-0.8em\TeX}}}

\setcopyright{acmcopyright}
\copyrightyear{2021}
\acmYear{2021}

\acmJournal{JOCCH}
\acmVolume{-}
\acmNumber{-}

\usepackage{paralist}
\usepackage{fancyvrb}

\long\def\comment#1{}
\newcommand{\q}[1]{\lq\lq{}{}#1\rq\rq{}{}}
\newcommand{\sq}[1]{\lq{}#1\rq{}}

\newcommand{\tool}[0]{FAST~CAT}
\newcommand{\toolteam}[0]{FAST~CAT~TEAM}

\begin{document}

\title{FAST CAT: Collaborative Data Entry and Curation for Semantic Interoperability in Digital Humanities}

\author{Pavlos Fafalios}
\email{fafalios@ics.forth.gr}
\orcid{0000-0003-2788-526X}

\author{Kostas Petrakis}
\email{cpetrakis@ics.forth.gr}

\author{Georgios Samaritakis}
\email{samarita@ics.forth.gr}

\author{Korina Doerr}
\email{korina@ics.forth.gr}

\author{Athina Kritsotaki}
\email{athinak@ics.forth.gr}

\author{Yannis Tzitzikas}
\authornote{Also affiliated with \textit{Computer Science Department, University of Crete, Heraklion, Greece}}
\email{tzitzik@ics.forth.gr}

\author{Martin Doerr}
\email{martin@ics.forth.gr}

\affiliation{%
  \institution{Centre for Cultural Informatics and Information Systems Laboratory, FORTH-ICS}
  \streetaddress{N. Plastira 100}
  \city{Heraklion}
  \country{Greece}
  \postcode{GR-70013}
}

\begin{abstract}
Descriptive and empirical sciences, such as History, are the sciences that collect, observe and describe phenomena in order to explain them and draw interpretative conclusions about influences, driving forces and impacts under given circumstances. Spreadsheet software and relational database management systems are still the dominant tools for quantitative analysis and overall data management in these these sciences, allowing researchers to directly analyse the gathered data and perform scholarly interpretation. However, this current practice has a set of limitations, including the high dependency of the collected data on the initial research hypothesis, usually useless for other research, the lack of representation of the details from which the registered relations are inferred, and the difficulty to revisit the original data sources for verification, corrections or improvements. 
To cope with these problems, in this paper we present \tool, a collaborative system for assistive data entry and curation in Digital Humanities and similar forms of empirical research. We describe the related challenges, the overall methodology we follow for supporting semantic interoperability, and discuss the use of \tool\ in the context of a European (ERC) project of Maritime History, called SeaLiT, which examines economic, social and demographic impacts of the introduction of steamboats in the Mediterranean area between the 1850s and the 1920s.
\end{abstract}

\ccsdesc[500]{Information systems~Information integration}
\ccsdesc[500]{Information systems~Data cleaning}
\ccsdesc[500]{Information systems~Mediators and data integration}
\ccsdesc[500]{Information systems~Extraction, transformation and loading}

\keywords{Data Entry, Data Curation, Semantic Interoperability, Archival Research, Digital Humanities}

\maketitle

\input{FastCat_Content}

\bibliographystyle{ACM-Reference-Format}
\bibliography{FastCat__BIB}

\end{document}

%% file: FastCat_Content.tex
\section{Introduction}

A vast area of research in descriptive and empirical sciences concerns the quantitative analysis of empirical facts, their description and interpretation of possible causes, influences and evolution trends.
Research in this case typically starts with the definition of the kind and formal structure of the data that needs to be extracted and/or transcribed from one or more data sources, for evaluation and further analysis of aggregated facts.  
This type of research is noticed in a wide range of disciplines, including History (e.g., analysis of historical archival sources~\cite{delis2020seafaring,petrakis2021}; our use case in this paper), Archaeology/Anthropology  (e.g., isotope or DNA analysis of ancient bone remains found in a particular area~\cite{haak2008ancient}), or Biodiversity (e.g., analysis of data from mesocosm experiments~\cite{pitta2017saharan}). 

However, current practice of quantitative analysis in these sciences nearly exclusively uses spreadsheets or simple relational databases, in order to organise the data as rows with multiple columns of related parameters.
This form offers itself for direct quantitative analysis under varying parameters, which may consequently be used for the scholarly interpretation of causes and impacts. 
However, it also has some typical problems including: 
i)~the high dependency of the digitised/transcribed data on the initial research hypothesis, usually useless for other research, 
ii)~the difficulty of representing nested tabular structures, 
iii)~the lack of representation of the details from which the registered relations are inferred,  
iv)~the difficulty to revisit the original sources of transcribed facts for verification, corrections or improvements, and
v)~the difficulty to combine and integrate data from multiple and diverse data sources, and thus perform (semi-)automated quantitative analysis of facts coming from different information sources \cite{doerr2008dream}.

In this paper, we focus on research workflows that require some kind of quantitative analysis of large amounts of data, such as transcripts of archival sources, and study the aforementioned problems in the context of the SeaLiT project, a European (ERC) project of Maritime History which examines economic, social and demographic issues of Mediterranean History between the 1850s and the 1920s emerging from the introduction of steamboats. 
Our objective is to support historians of this project in digitising and curating their unique information sources collaboratively, for enabling long-term preservation, semantic interoperability, and the use of the data as a primary source for current and future historical research.

To this end, we first detail the data management requirements, the respective challenges, as well as the overall methodology we follow for supporting semantic interoperability. 
We then describe \tool, a collaborative system that we have developed for data entry and consolidation. We present its user interface, the supported  functionalities, its configurability, as well as its use by a large number of historians within the SeaLiT project.  

The rest of this paper is organised as follows:
Section~\ref{sec:Context}
discusses the context of this work and the 
related requirements and challenges.
Section~\ref{sec:approach}
describes the overall methodology we follow for data entry, data consolidation and data exploration, and discusses related works.  
Section~\ref{sec:fastcat} describes
the \tool\ tool, its configurability, its data entry interface, as well as the \toolteam\ environment for data curation. 
Section~\ref{sec:usecase} describes the use of \tool\ in the context of the SeaLiT project.
Finally, Section~\ref{sec:conclusion} concludes the paper and discusses future work.

\section{Context, Requirements and Challenges}
\label{sec:Context}

\subsection{Context: The SeaLiT Project}
\label{sec:Sealit}

SeaLiT\footnote{SeaLiT - \textit{Seafaring Lives in Transition. Mediterranean Maritime Labour and Shipping during Globalisation, 1850s-1920s}. ERC Starting Grant - ID: 714437. \url{http://www.sealitproject.eu/}} 
is an ongoing research project of  Maritime History that explores the transition from sail to steam navigation and its effects on seafaring populations in the Mediterranean and the Black Sea between the 1850s and the 1920s. Historians and researchers in this project investigate, besides others, the maritime labour market, the evolving relations among ship-owners, captain, crew and local societies, and the development of new business strategies, trade routes and navigation patterns, during the transitional period from sail to steam. 

The information management challenge that SeaLiT faces is the ability to faithfully catalogue historical data sources for use as a primary source for research, while integrating the data into a common form from which historical analysis and questions can be carried out efficiently.
The information sources used in SeaLiT range from handwritten \textit{ship logbooks}, \textit{crew lists}, \textit{payrolls} and \textit{student registers}, to \textit{civil registers}, \textit{business records}, \textit{account books} and \textit{consulate reports}, gathered from different authorities (i.e., not following a common format) and written in different languages (Spanish, Italian, French, Russian, Greek). 

\subsection{Data Management Requirements and Challenges}
\label{sec:RC}

The data management requirements can be grouped into three categories: 
\begin{itemize}
    \item Data transcription
    \item Data consolidation
    \item Data analysis and exploration
\end{itemize}

Data transcription (or digitisation) refers to the ability of the researchers to faithfully catalogue their information sources, aiming at long-term preservation and exploitation of the data beyond the objectives of a particular research activity. Here the main challenges include: 
i)~how to transcribe/catalogue as much relevant information as possible and as exact as possible, 
ii)~how to support fast data entry, and 
iii)~how to handle uncertainty in the original data. With respect to the latter, a common problem in SeaLiT is the difficulty in recognising one or more characters in some text of the original source. In some cases the historian can make a guess, while in other cases this is impossible. 

Data consolidation refers to the integration of data coming from different sources into a common form from which analysis and questions can be carried out efficiently. It also involves the correction or normalisation of values and the disambiguation of entity identities, like the indication that two different person instances (written in different ways or with spelling errors) are actually the same person. 
Typically known as \textit{instance matching}, we stress in this context the managing of varying identity assumptions.
Here the main challenge is how to integrate and curate the digitised information of the different data sources and in parallel keep track of where the information comes from and what its original form is. A particular technical challenge is how to handle identities and support a multi-level instance matching process (within the same document, across related documents, across independent documents), each time with different default assumptions and background knowledge that combines automation and manual intervention (where \lq{}identity\rq{} is understood as what the author meant writing the expression at hand). Another challenge is how to handle vocabularies or thesauri of terms, i.e., universals in contrast to instances, especially in case of multilingual data.
The key success factor is the transformation of the data to a single, curated, connected and provenance-aware resource that can be queried and analysed efficiently. 

Data analysis and exploration refers to the exploitation of the consolidated data, e.g. for quantitative analysis of collective phenomena for drawing conclusions on possible impact factors. Researchers need to be able to find answers to complex information needs that require combining complementary information from different sources. The challenge here is how to provide users with user-friendly interfaces that support them in finding the needed information easily and in different forms, like tables or charts, for direct use in their research. Thus, the key success factors are usability as well as trustworthiness which can be achieved by providing access to the original sources (e.g., for validation).

Apart from the aforementioned technical challenges, there are some common organisational-related challenges. At first, in large research projects there are usually research teams that are spread across the world and which need a common and secure place for storing and accessing their data, sharing it internally, and releasing parts of it to a wider audience when they want to do so. This requires a plan and tools for collaborative work. Moreover, there might be inconsistent access to Internet during particular research activity, like field studies or working in remote, local archives. This might require offline data entry and synchronisation when Internet is available. Finally, in ongoing research projects where new data sources might become available at any time, there is the need of a clear strategy on how to handle updates. New data sources means new structures for data entry and storage, new entity instances that require instance matching, new vocabulary terms, etc. 

In this paper, our focus is on \textit{data transcription} and \textit{data consolidation}, while we also try to cope with the organisational-related challenges.

\section{Overall Approach and Related Work}
\label{sec:approach}

The focus of the overall approach we follow for coping with the challenges described in the previous section is \textit{semantic interoperability} \cite{ouksel1999semantic}. 
This notion is defined as the ability of computer systems to exchange data with unambiguous, shared meaning. 
This can be achieved by adding metadata about the data and linking each data element to an ontology or controlled/shared vocabulary. 
Semantic interoperability is a requirement to enable, amongst others, knowledge discovery, inference and data federation between information systems, in particular when data sources from different providers need to be combined.

To this end, the semantic artefacts that we consider for allowing both humans and machines to locate, access and understand the data and metadata are the following:
\begin{itemize}
    \item  A core domain ontology, in particular the ISO standard CIDOC-CRM (more below), used for representing all entities and entity associations that appear in the data.
    \item A set of controlled vocabularies (thesauri of terms), represented through the SKOS W3C recommendation\footnote{\url{https://www.w3.org/TR/skos-reference/}}; the vocabularies are used for controlling the data entry process and ensuring uniformity and consistency in the storage and retrieval of the different notions that appear in the data.
    \item A configurable URI generation policy applied during the data transformation process for assigning unique (Web) identifiers to the entities that appear in the data. 
\end{itemize}

\begin{figure}[h]
	\centering
	\includegraphics[width=0.65\textwidth]{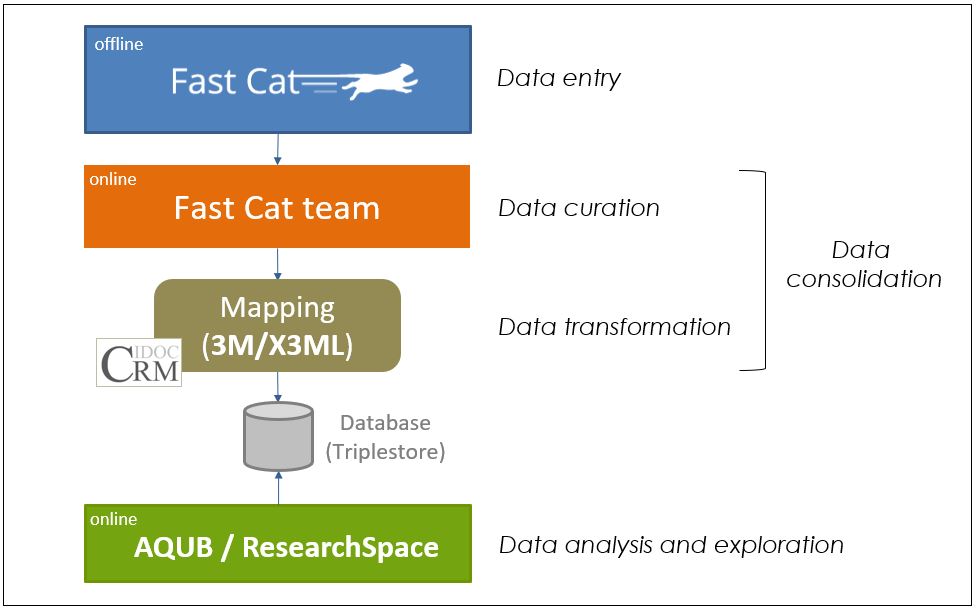}
	\caption{The overall data management approach}
	\label{fig:pipeline}
\end{figure}

Fig.~\ref{fig:pipeline} depicts the overall data management methodology we follow for supporting semantic interoperability in Digital Humanities and similar forms of empirical research.
There are four main steps in this methodology, each one supported by a software system: 
i)~data entry (supported by \tool), 
ii)~data curation (supported by \toolteam),
iii)~data transformation (supported by X3ML and 3M Editor~\cite{marketakis2017x3ml}), and  
iv)~data analysis and exploration (supported by tools like AQUB~\cite{kritsotakis2018assistive} or ResearchSpace~\cite{oldman2018reshaping}).

Below, we first detail each of these steps (Sections \ref{subsec:apprDataEntry}-\ref{subsec:dataAnalysisExpl}) and then discuss related works (Section \ref{subsec:rw}).

\subsection{Data Entry}
\label{subsec:apprDataEntry}
The first step is \textit{data entry} which is supported by the \tool\ system (detailed in the next section). This can be done by a user either online, or offline with possibility of automated synchronisation when the user gets online. 

In  \tool, data from different information sources can be transcribed as \lq{}records\rq{} belonging to specific \lq{}templates\rq{}, where a \lq{}template\rq{} represents the structure of a single data source. 
A \lq{}record\rq{} in \tool\ organises the data and metadata in tabular form (tables), similar to spreadsheets, offering functionalities like nesting tables and selection of term from a controlled vocabulary (more below).  
Here researchers can agree on the patterns to use when there is some kind of uncertainty in the original data. For instance, historians in SeaLiT put a question mark when a specific character in the original archival source is not recognisable, or put the text inside square brackets when they are not sure about a set of characters. 

The advantage of using tables for data entry is twofold. First, historians are familiar with this form due to the wide use of spreadsheet programs like Microsoft Excel. Second, the tables facilitate fast data entry and error detection because the process is iterative and the user can easily inspect the entries of the previously-filled rows.

\subsection{Data Curation}
\label{subsec:apprDataCuration}
The second step is \textit{data curation} supported through \toolteam, an environment offered by \tool\ that allows the collaborative curation of the transcribed data through the management of \textit{entities} and \textit{vocabularies (thesauri of terms)}. The reason of decoupling data curation from data entry, and avoiding curating the data directly in the records, is the need to maintain the transcribed data as close to the original information sources as possible. 
This is of crucial importance for researchers since the data consolidation may be ambiguous and require further research and repeated revision at any time in the future. It further enables them to consolidate the data for exploitation in other services (like the data exploration services that we will see below) without having to change the information as it appears in the archival sources (and thus \sq{spoil} the data). This also means that, other researchers working in some other project can work with the original transcribed data and apply their own consolidation processes.

With respect to the management of \textit{entities}, researchers can inspect the main entity instances that appear in the data (names of \textit{persons}, \textit{locations}, \textit{ships}, and \textit{legal entities}, in the case of SeaLiT) and start curating them. Here a first automated curation step considers a set of rules for giving the same identity to a set of entity instances having some common characteristics/properties.
For example, all person instances in the same source that have the same \textit{firstname}, \textit{lastname}, \textit{father's name} and \textit{birth date} must be considered as the same person and thus have the same identity.  
Then, the available curation actions include: 
i) corrections of entity names or other entity properties (e.g., changing \q{Vodowice} to \q{Wodowice}, or \q{G??rge} to \q{George}), 
ii) indication that two or more entity instances refer to the same real-world entity, thus they must have the same \lq{}identity\rq{} (manual instance matching process), and
iii) indication that a specific instance from a set of automatically matched instances is a different entity and  thus must have a different \lq{}identity\rq{}. 
Note here that, when making corrections on the entity names or properties, the transcribed data in the records does not change. All entity instances are linked to their occurrences in the transcripts. This means that researchers can \sq{go back} at any time, visit the original transcribed data, and inspect how exactly a particular entity appears in the original data sources. 
As we described previously, this is very important for validation purposes and for keeping track of provenance information. 

With respect to the management of \textit{vocabularies}, researchers can maintain lists of terms in different languages, thus allowing for cross-lingual data exploration. Specifically, for each term in a vocabulary, users can provide a preferred term in English as well as its broader term (if any). 
The storage of broader terms provides an hierarchy for the terms, which can be very useful when exploring the data. For example, one can retrieve all data related to a general term through its narrow terms.

The curated data derived by this step, together with the original transcribed data, can be directly exploited by other applications before their transformation. For example, the work in~\cite{petrakis2021} shows how the data of \textit{ship logbooks} of the nineteenth and twentieth centuries, digitised and curated using \tool, are visualised on a web-accessible map application called \textit{Ship Voyages}.\footnote{\url{http://www.sealitproject.eu/digital-seafaring}}

\subsection{Data Transformation}
\label{subsec:apprDataTransformation}
The third step in our data management pipeline is \textit{data transformation}. Here the objective is to model and semantically represent the (curated) data and metadata using established data models, for supporting data exchange, interoperability and long-term validity, and make the data exploitable beyond a particular research problem or project. 

To achieve this, we first need to decide on the \textit{domain ontology} to use for representing and integrating the data of all FAST CAT records. 
Here we choose to create a model that is compatible with the CIDOC Conceptual Reference Model (CRM)\footnote{\url{http://www.cidoc-crm.org/}} \cite{doerr2003cidoc}. 
CIDOC-CRM is a high-level, event-centric ontology (ISO standard\footnote{\url{https://www.iso.org/standard/57832.html}}) of human activity, things and events happening in spacetime, providing definitions and a formal structure for describing the implicit and explicit concepts and relationships used in cultural heritage documentation and beyond. 

Given the domain ontology, we then have to create the \textit{schema mappings} that map the elements of a template (e.g., a table column) to classes and properties of the domain ontology. This is a time-consuming process that need to be done for each different template and which may require many revisions as long as the data engineer better understands the data or changes are made to the FAST CAT templates.
This process is supported by the X3ML mapping framework and its user interface 3M Editor \cite{marketakis2017x3ml}. The framework allows data engineers to relate equivalent concepts or relationships from the source schemata (\tool\ templates, in our case) to a target schema (the domain ontology), as well as to define the URI generation policies that assign identifiers to the entities that appear in the data. 

After having defined the schema mappings for each different \tool\ template as well as each category of entities in \toolteam, the data can be automatically transformed (through X3ML) to a rich semantic network of Linked Data (RDF triples)  \cite{heath2011linked} and be ingested in a semantic repository (RDF database/triplestore), from where data analysis and exploration can be initiated.
Whereas data columns in the transcripts basically appear associated in the order next to each other as in the original sources, possibly in nested tables, the transformed data form a connected semantic network that completely connects fields in the data columns with meaningful relationships (RDF properties) and other implicitly inferred entities.

In Section~\ref{sec:usecase} we provide an example of how we model maritime history data in the context of SeaLiT using a CIDOC-CRM compatible domain ontology, as well as an example of a schema mapping and the corresponding transformed data. Since the definition of schema mappings and the data transformation process is out of the score of this paper, we suggest the reader to refer to \cite{marketakis2017x3ml} for further information.

\subsection{Data Analysis and Exploration}
\label{subsec:dataAnalysisExpl}
The user-friendly exploration of a semantic repository (containing the semantic network derived from the data transformation process) can be performed through two main general access methods: 
i)~\textit{keyword search}, where the user submits a free text query and gets back a ranked list of results that are relevant to the query terms (e.g.,~\cite{kadilierakis2020keyword,elas4rdf,nikas2020keyword}), and
ii)~\textit{interactive access}, where the user explores the data through intuitive interactions with a data access system, e.g., using a \textit{faceted search} interface  \cite{tzitzikas2017faceted} or \textit{assistive query building} (e.g.,  \cite{kritsotakis2018assistive}).  

In SeaLiT we make use of \textit{ResearchSpace}~\cite{oldman2018reshaping}, an open source platform build on top of the \textit{metaphactory} platform~\cite{haase2019metaphactory}. The platform offers a variety of functionalities, including an \textit{assistive query building} interface that allows users to gradually build complex queries and analyse the results quantitatively through different visualisations. Such queries actually follow paths in the semantic network, defined through one or more relationships between the main entities that appear in the data. 

\begin{figure}[h]
	\centering
	\includegraphics[width=\linewidth]{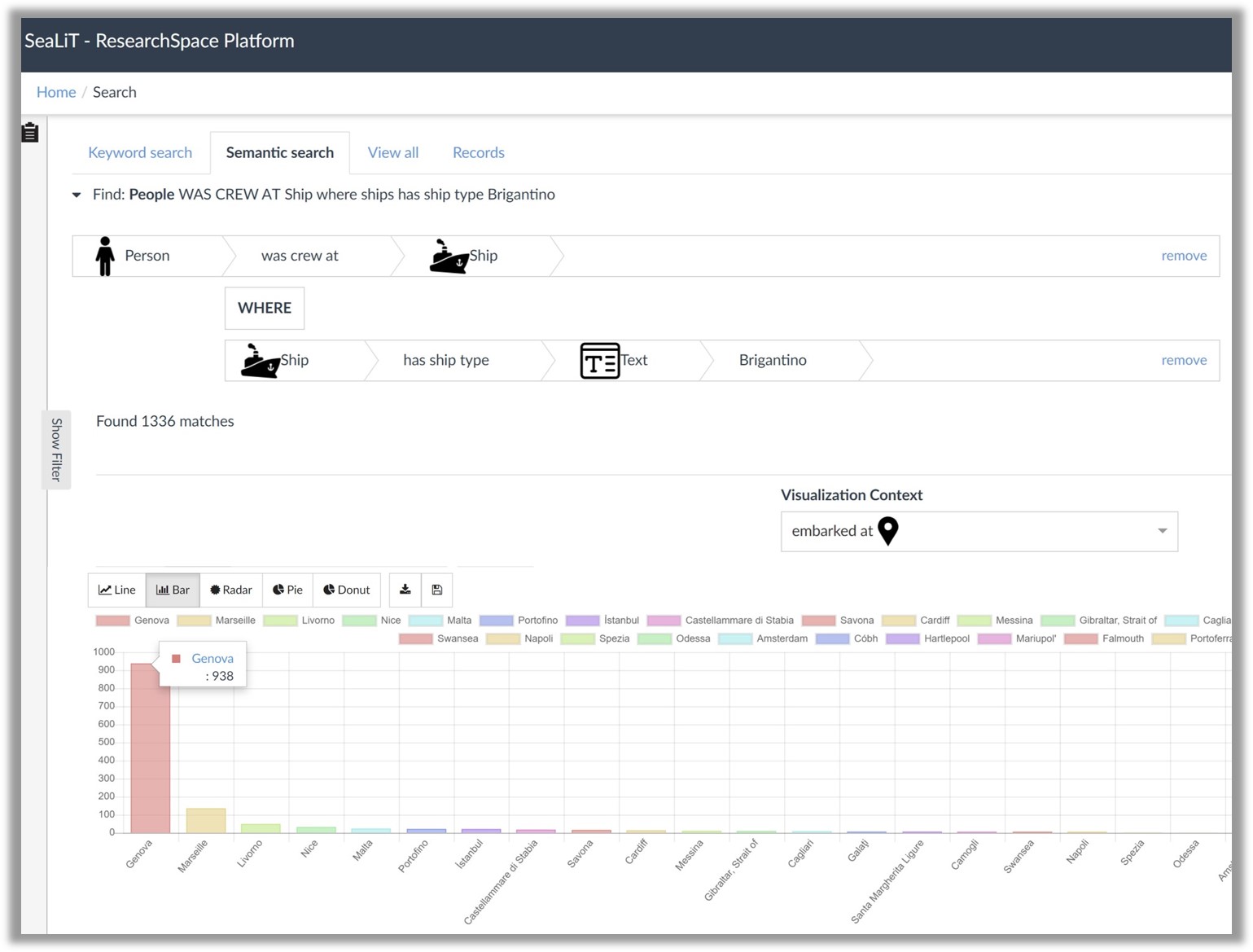}
	\caption{Data analysis and exploration in the \textit{ResearchSpace} platform.}
	\label{fig:researchSpaceExample}
\end{figure}

Fig.~\ref{fig:researchSpaceExample} shows a screen dump of the system, where the user inspects a bar chart showing the \textit{embarkation location of persons that were crew members at ships of type brigantine}. 
In the upper part of the figure we see the query built by the user (Person - was crew at - Ship, where Ship - has ship type - Brigantino),  while in the lower part the user has selected to view a bar chart of the results using the visualisation context  (aggregation function) \sq{embarked at}. 

\subsection{Related Work}
\label{subsec:rw}

Since the focus of our work is data entry and consolidation for semantic interoperability and the production of a semantic network of Linked Data \cite{heath2011linked}, below we discuss works that cover this analysis pipeline.

At first, the original data of interest may be stored and managed in a wide range of formats, e.g., spreadsheets, relational databases, and/or textual documents (unstructured or semi-structured). In descriptive and empirical sciences, like History, spreadsheet software (like Microsoft Excel or Google Sheets) and relational database management systems (RDBMS), like Microsoft Access or MySQL, are still the dominant tools for data entry, storage and management. 

With respect to RDBMS-based systems, HEURIST\footnote{\url{http://heuristnetwork.org/}} is an open source  platform that allows Humanities researchers to design, manage, analyse and publish their own MySQL databases through a user-friendly web interface. HEURIST offers a wide range of functionalities for the management of research data, including the ability to import existing datasets, upload media files, create web pages, as well as manage sophisticated searches.
In the same context, nodegoat\footnote{\url{https://nodegoat.net/}} is a data management tool that allows scholars to build datasets based on their own data model, offering relational modes of data analysis. 
The system follows an object-oriented approach treating people, events, artefacts and sources as objects that are connected through relations. Each object can be supplemented with geographical and temporal attributes, thus allowing geographic and diachronic visualisations. Nodegoat is also capable of processing complex queries by means of filtering and analysis over the networks formed by the relations between the objects.
Our data management methodology and the \tool\ system described in this paper follow a much different approach compared to the above systems. Our approach is oriented towards semantic interoperability, so that both humans and machines can locate, access and understand the data and metadata.

Data curation includes all the processes needed for controlled data maintenance and management, together with the capacity to add value to data \cite{10.5555/2726970.2726972}. 
From a theoretical point of view, the white paper in \cite{munoz2011issues} provides background and provokes discussions about the skills, professional roles, training, and institutional support needed for data curation in humanities research. \cite{palmer2013foundations} describes an approach to data curation education that integrates traditional notions of curation with principles and expertise from library, archival, and computer science. \cite{henry2014data} describes the life cycle of data curation for the Humanities, including discussions about staffing, infrastructure, roles, responsibilities, documentation as well as interesting cases studies. 

From a technical perspective, the Data Tamer \cite{stonebraker2013data} system, and its commercial realisation Tamr\footnote{\url{https://www.tamr.com/}}, is an example of an end-to-end data curation system, offering components for schema integration, entity consolidation, data visualisation as well as crowd-sourcing. 
An important sub-process in data curation is \textit{data cleaning}, which deals with detecting and removing errors and inconsistencies from data in order to improve their quality \cite{rahm2000data}. 
\cite{chu2016data} presents a taxonomy of the data cleaning literature, describes the state-of-the-art techniques, and also highlights their limitations.
OpenRefine\footnote{\url{https://openrefine.org/}} \cite{verborgh2013using} is a popular desktop application for data cleaning, also supporting the transformation of the data to other formats. It operates on rows of data which have cells under columns, which is very similar to relational database tables. \cite{krishnan2016towards} presents the results of an interesting user survey of 29 data analysts and infrastructure engineers about data cleaning processes. The results highlight three important themes: i)~the non-linear and iterative nature of data cleaning, ii)~the lack of rigour in evaluating the correctness of data cleaning, and iii)~the disconnect between the analysts who query the data and the infrastructure engineers who design the cleaning pipelines.
Contrary to existing works on data curation that mainly focus on correcting errors and inconsistencies in the data, in \tool\ we  follow a \textit{provenance-aware} approach that also incorporates \textit{instance matching} processes and the management of \textit{vocabularies of terms}. 

With respect to the final step of transforming the data sources to Linked Data, there is a plethora of solutions to execute mappings from different file structures and serialisations to RDF. R2RML\footnote{\url{https://www.w3.org/TR/r2rml/}} is a W3C standard for mapping \textit{relational databases} into RDF, while  \cite{dimou2014rml} describes the RML mapping language, an  extension of R2RML for mapping \textit{heterogeneous sources} into RDF. \cite{michel2014survey} surveys existing approaches on mapping relational databases to RDF, analyse their commonalities and differences, and underline common patterns in the orchestration of solutions applicable to each step of the translation process.
X3ML \cite{marketakis2017x3ml} is an XML-based mapping language and framework for the management of the core processes needed to create, maintain and manage mapping relationships between different data sources. It allows the description of both \textit{schema mappings} and \textit{URI generation policies}, and offers tools (like the 3M Editor) for managing, editing, visualising and executing those mappings. The same work also provides a detailed overview and categorisation of works on mapping heterogeneous data sources to RDF. 
In the same context, MatWare \cite{tzitzikas2014matware} is a framework that automates the construction and update of a \textit{semantic warehouse} (RDF database), by implementing a pipeline that integrates data from multiple different sources, also offering services for instance matching and quality assessment.
In our work, we make use of the X3ML framework for mapping the curated data to the domain (target) ontology. 

As regards the domain of historical research (our use case), the survey in \cite{merono2015semantic} studies the joint work of historians and computer scientists in the use of Semantic Web methods and technologies as well as how such works help in solving open problems in historical research. The survey provides an extensive analysis on works and systems for i)~knowledge modelling (historical ontologies), ii)~text processing and mining, iii)~search and retrieval, and iv)~data integration for semantic interoperability, discussing also aspects of the Semantic Web that could be furtherly exploited in historical research.
Two such aspects are \textit{\q{non-destructive data transformations}} and \textit{\q{linking more historical data}}. 
The high provenance-awareness of our approach (which decouples data curation from data entry; as described in Section \ref{subsec:apprDataCuration}), together with the use of standard models for representing and publishing the data in the \tool\ records (CIDOC-CRM, SKOS), are characteristics of our approach that help towards this direction. 

In the same context of historical research, \cite{koho2019warsampo} presents the WarSampo knowledge graph, a
shared semantic infrastructure for publishing data about WW2 and Finnish
military history as Linked Data. The shared semantic infrastructure is based on the idea of representing war as a spatio-temporal sequence of events that soldiers, military units, and other actors participate in. The considered metadata schema is an extension of CIDOC-CRM, supplemented by various military history domain ontologies. In addition, to support sustainability, a repeatable automatic data transformation and linking pipeline has been created that rebuilds the whole semantic network from the individual source datasets. Note here that, contrary to our work which focuses on data entry and curation (performed by the researchers), WarSampo is a pure data transformation infrastructure for historical datasets. 
Other semantic databases of Finnish history are accessible through the Linked Data Finland research initiative.\footnote{\url{https://www.ldf.fi/datasets.html}}

\section{The \tool\ Tool}
\label{sec:fastcat}

Below we detail the functionality and the user interface offered by \tool. We first describe how we can configure \textit{templates} of records, each one representing a different data source (Section~\ref{subsec:config}). Then, we show how one can create records and start filling the data entry forms (Section~\ref{subsec:dataentry}), and describe the data curation functionality offered by \toolteam\ (Section~\ref{subsec:datacuration}). Finally, provide some technical implementation details (Section~\ref{subsec:techDetails}).

\subsection{Configuration of Templates}
\label{subsec:config}

For initiating the data entry process, we first need to construct the \textit{templates}, i.e., the data entry forms used for transcribing the data of a particular source. A template actually represents the structure of a data source and allows the documentation of metadata information about both the source and the data entry process (like who makes the transcription). Given a template, the user can then create a \textit{record} and start transcribing the source data. 

\begin{figure}[h]
	\centering
	\includegraphics[width=\linewidth]{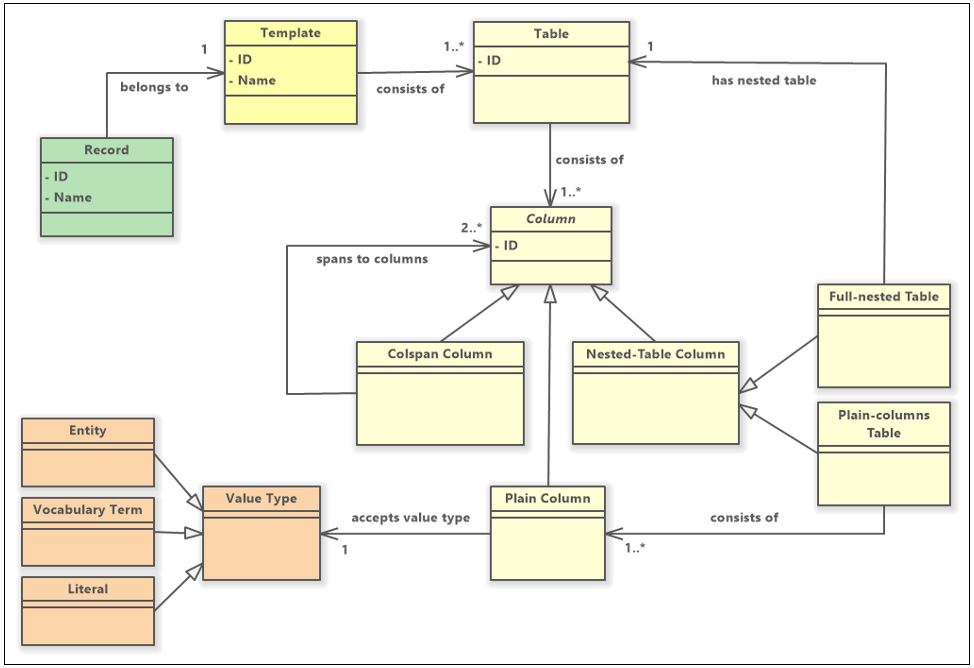}
	\caption{Class diagram describing the configuration model in \tool.}
	\label{fig:classDiagram}
\end{figure}

Fig. \ref{fig:classDiagram} depicts a class diagram of the main notions involved in the configuration of a template. 
At first, a \textit{template} consists of one or more \textit{tables} and a \textit{table} of one or more \textit{columns}. 
A column is an abstract class with three different specialisations:
i) \textit{colspan column}, 
ii) \textit{plain column},
iii) \textit{nested-table column}.
A \textit{colspan column} is a table column which does not accept values but just organises a set of other columns. For example, PERSON is a colspan column which groups together the columns: FIRST NAME, LAST NAME, FATHER'S NAME, BIRTH DATE. 
A \textit{plain column} is a table column which accepts values, i.e., the user can provide a value while filling a table row, like writing some text, or selecting a term from a controlled vocabulary.
A \textit{nested-table column} is a table column which spans to another table, so the user can link the value of this column to a whole table of multiple rows. For instance, a table whose each row describes information for a particular person can contain a nested-table column with name CHILDREN which allows filling information (like FIRSTNAME, LASTNAME and BIRTH DATE) for all person's children. 

A \textit{nested-table column} can be implemented in different ways, depending on the functionality and flexibility we want to offer to the end-users. \tool\  currently supports two implementations of a nested-table column: 
i)~\textit{plain-columns table}, and 
ii)~\textit{full-nested table}. 
A \textit{plain-columns table} is a simple table consisting of one or more \textit{plain columns} (so the user can provide multiple values organised in plain columns), while a \textit{full-nested table} is a table consisting of columns of any type.
The former type is applicable when the user needs to provide multiple values to one or more columns within a table row (e.g., multiple ship owners). The latter type is applicable for cases where we want to add additional information about a table record (i.e., an entire table row) and this information requires multiple columns (e.g., information about all students enrolled in a course). 
Examples of the different column types are shown in Figure \ref{fig:colExamples}.

A plain column accepts values of different types, in particular: 
i)~Entity (the value is the name or property of an entity, e.g., of a person),
ii)~Vocabulary Term (the value is a term from a controlled vocabulary), 
iii)~Literal (the value is a literal, e.g., a free text, a number, or a date). 
Each of these value types can contain additional information, like the type of literal in the case of \sq{Literal} (e.g., String, Integer, Date, etc.), or the type of entity in the case of \sq{Entity} (e.g., Person, Location, Ship, etc.).

\subsection{Data Entry}
\label{subsec:dataentry}

\begin{figure}[h]
	\centering
	\includegraphics[width=\linewidth]{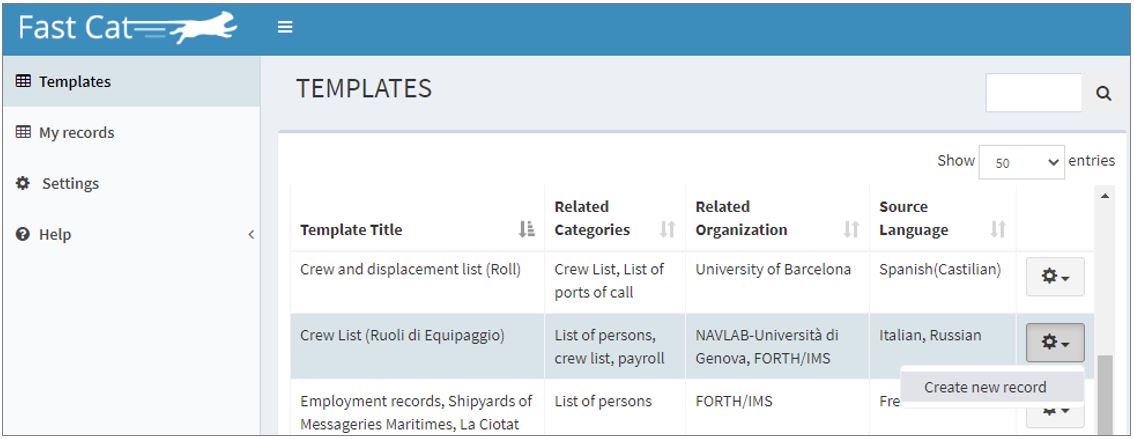}
	\caption{Available templates and creation of a new template record in \tool.}
	\label{fig:newRecord}
\end{figure}

Data entry is performed through \tool's web interface.
Its home page shows a table containing all the available templates together with some basic information for each template (Fig.~\ref{fig:newRecord}).
For initiating the data entry process, the user first needs to create a new \textit{record} for one of the available templates. 
After filling some basic record information, the record is created and the user can start filling the data entry forms. 
Whenever a user saves a record, \tool\ saves it locally and also (if online) remotely. In most cases, no conflicts arise since one particular record is usually managed by a single person. In case of a conflict, we handle it by checking the \q{Last modified} field and declaring the newer version as the final one, or allowing the user to set the correct version. 

\begin{figure}[h]
	\centering
	\includegraphics[width=\linewidth]{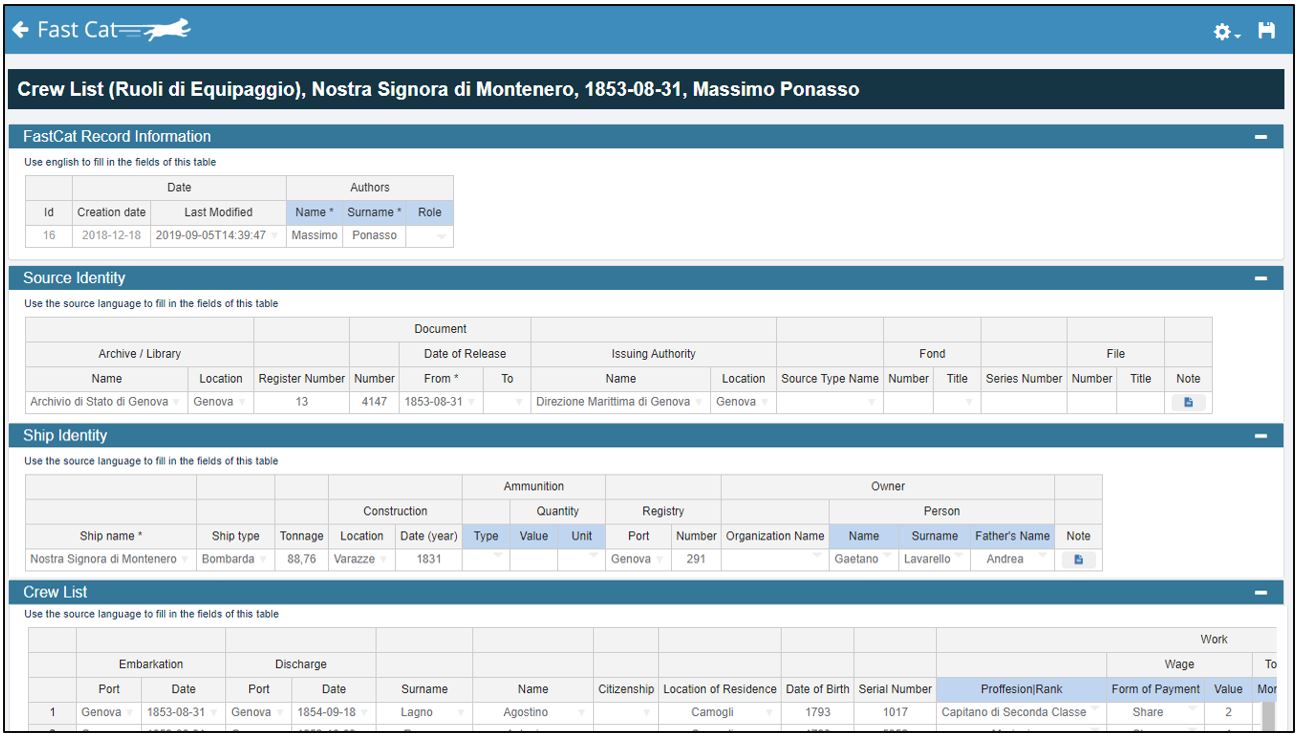}
	\caption{The data entry forms in a \tool\ record.}
	\label{fig:dataentry}
\end{figure}

\begin{figure}[h]
	\centering
	\includegraphics[width=\linewidth]{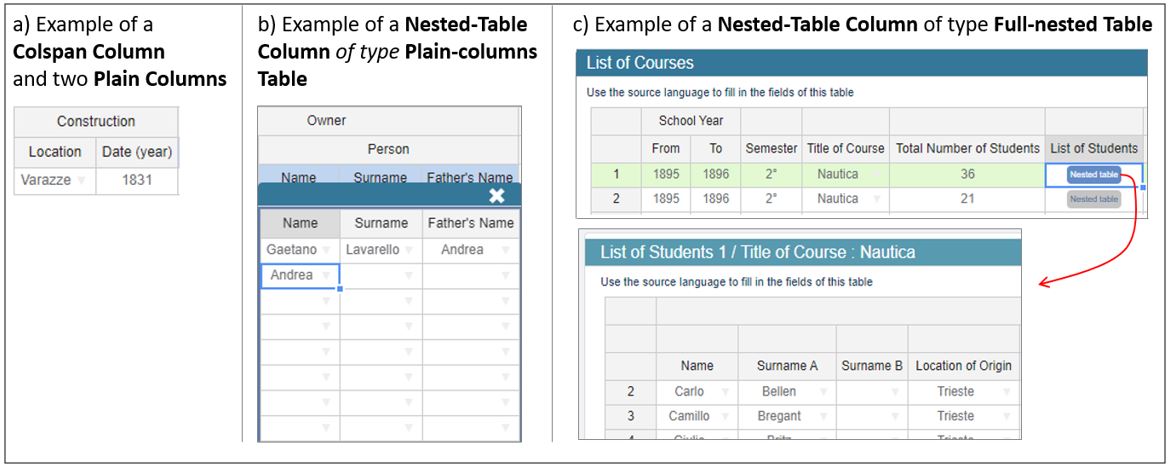}
	\caption{Examples of the different column types in \tool.}
	\label{fig:colExamples}
\end{figure}

Figure~\ref{fig:dataentry} shows an example of a record from the SeaLiT project (transcription of an archival \textit{Crew List}), while examples of the different column types (as described in Section \ref{subsec:config}) are shown in Fig.~\ref{fig:colExamples}. 
In a plain column (Fig.~\ref{fig:colExamples}-a), the user can either write some text, select a term from a controlled (predefined or dynamic) vocabulary of terms, or fill a date. 
In case of a nested-table column of type \textit{plain-columns table} (Fig.~\ref{fig:colExamples}-b), the corresponding columns are displayed in blue colour and data entry is performed through a \sq{pop-up} table below the columns. On the contrary, a \textit{full-nested table} is displayed in a new area below the first table (Fig.~\ref{fig:colExamples}-c).
Users are also able to export the record data to Excel or XML. This allows them to perform external analysis of (part of) the data, e.g., for creating charts in Excel.

\subsection{Data Curation using \toolteam}
\label{subsec:datacuration}

\begin{figure}[h]
	\centering
	\includegraphics[width=\linewidth]{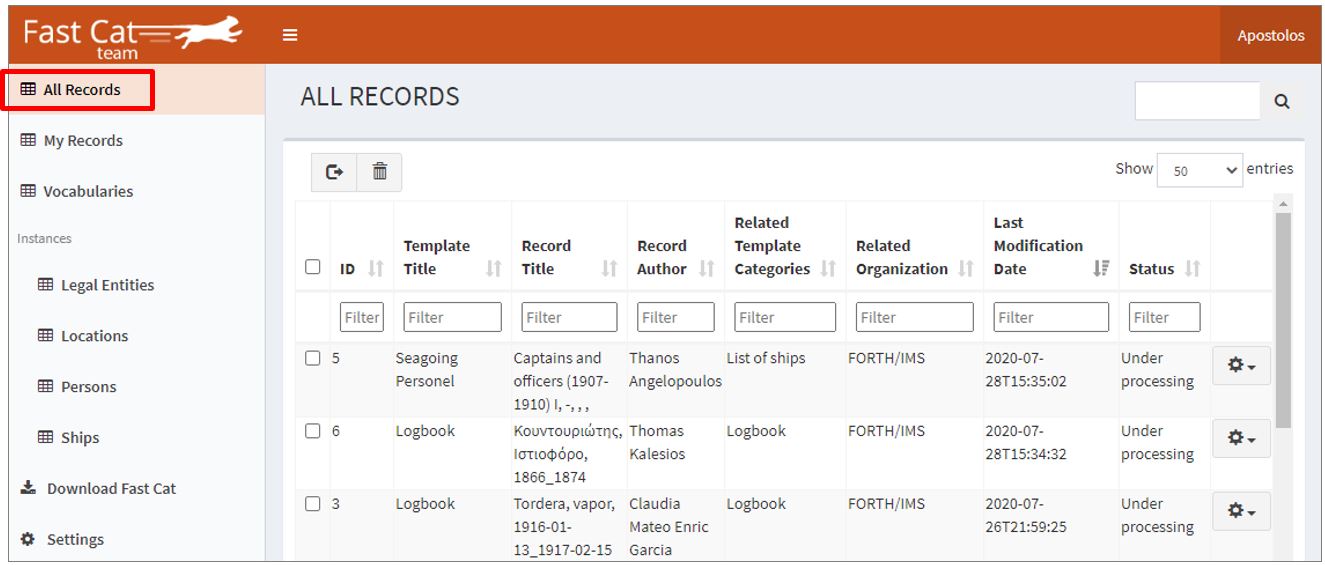}
	\caption{The home page of \toolteam.}
	\label{fig:fastcatteam}
\end{figure}

The collaborative curation of the transcribed data, in particular the \textit{vocabularies} and the \textit{entities}, is performed through a dedicated \tool\ environment called \toolteam. 
The home page of \toolteam\ shows a table of all publicly shared records (Fig.~\ref{fig:fastcatteam}). For each record, the table shows its ID and title, the corresponding template title, as well as some metadata information like the author of the record, its last modification date and its status. 
The status information can take one of the following values: 
\sq{Under processing} (data entry is still in process), \sq{Ready for review} (data entry has been finalised but review and/or curation is still in process), 
\sq{Reviewed and Ready for publishing} (data entry and curation has been finalised and the data is ready for publication), 
\sq{Published} (the data has been transformed and ingested in the data analysis and exploration service).
This workflow is not linear, i.e., the user can transition from one state to any other state. For example, the user can move from the \sq{published} state to the  \sq{under processing} state because, for instance, new data needs to be added in the corresponding record. Similarly, one can move from \sq{published} to \sq{Ready for review} because additional curation processes need to be performed. 

\begin{figure}[h]
	\centering
	\includegraphics[width=\linewidth]{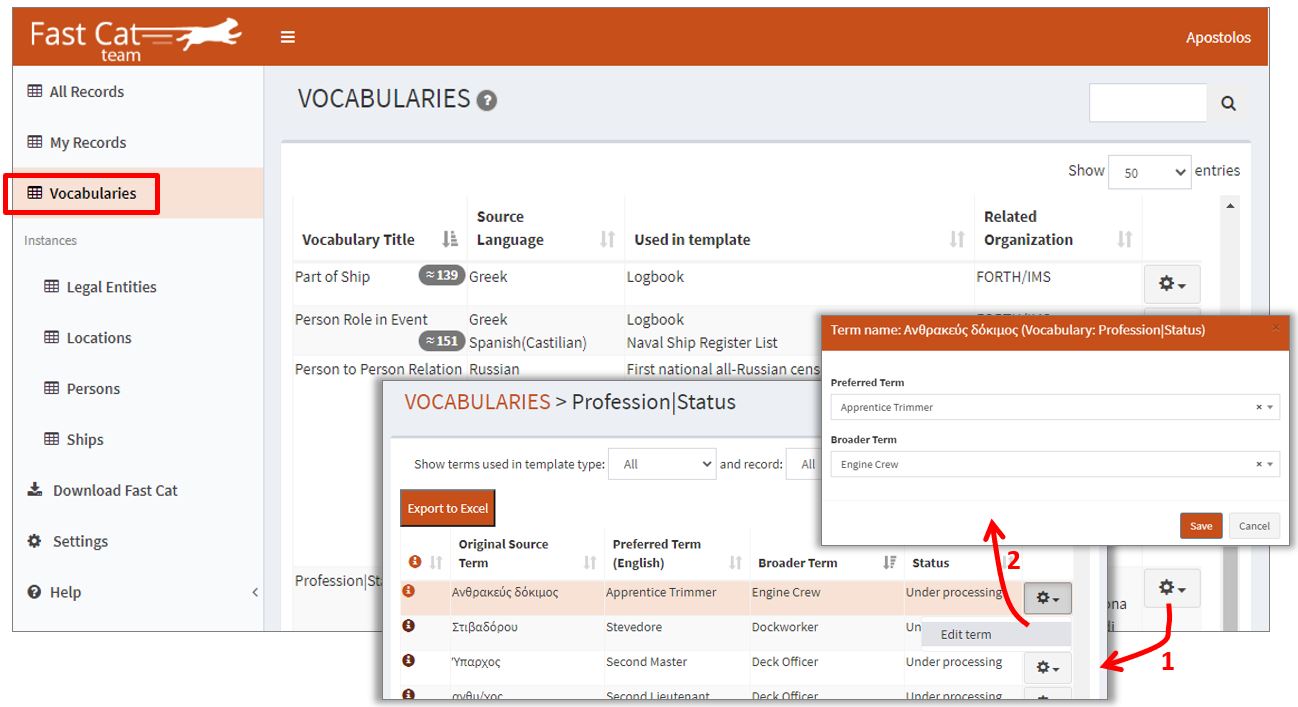}
	\caption{Management of vocabularies in \toolteam.}
	\label{fig:vocs}
\end{figure}

Through the \sq{Vocabularies} menu item, the user can see a table with all vocabularies used in the \tool\ records and start their curation (Fig.~\ref{fig:vocs}). The table shows the name of the vocabulary, the source language(s), the template(s) in which it is used, and the involved organisations (that take care of the curation). For each term in a vocabulary, the user can set a \textit{preferred term} in English and its \textit{broader term}.

\begin{figure}[h]
	\centering
	\includegraphics[width=\linewidth]{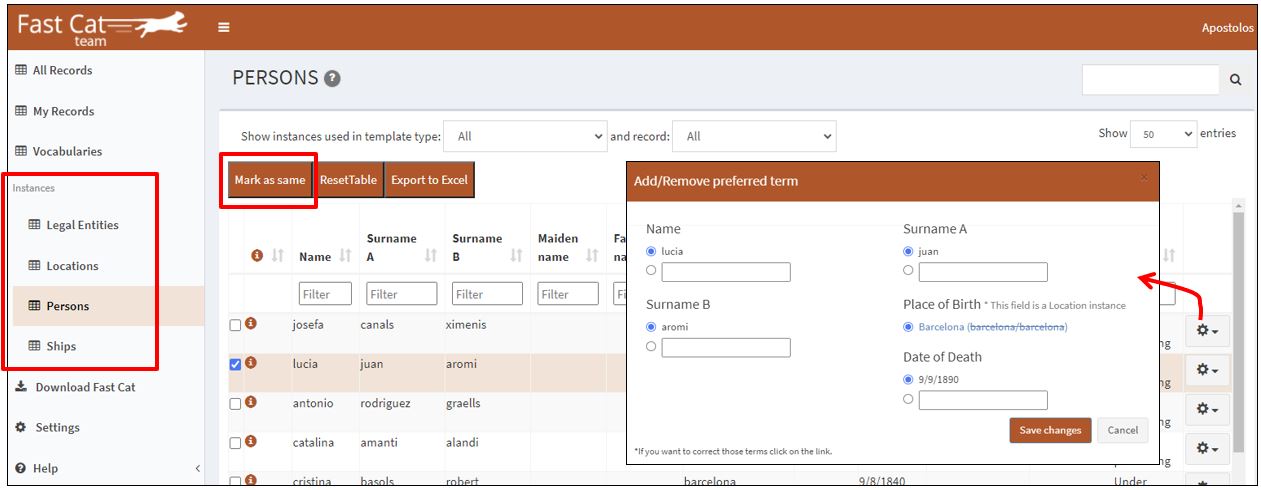}
	\caption{Management of entities in \toolteam.}
	\label{fig:instancesPers}
\end{figure}

For curating the \textit{entity instances}, \toolteam\ shows (in a left menu) a list with all the (predefined) entity types whose instances exist in the records (Fig.~\ref{fig:instancesPers}). In SeaLiT, for example, there are four entity types: \sq{Legal Entities}, \sq{Locations}, \sq{Persons}, and \sq{Ships}.
When selecting to start curating one of the entity types, the user is shown a table of all its instances. The information shown in the table is different for each type of entities. 
For persons, for example, the table shows the following information: name, surname(s), maiden name, father's name, place and date of birth, date of death, registration number, status/capacity/role.
The user can select an entity and change/set information, or select two or more entities and indicate that they correspond to the same real-world entity (\textit{manual instance matching}). In the latter case, the user must also indicate the preferred value for the entity properties (if there is a conflict). 
Note here that, any change in the values do not alter the original values as stored in the \tool\ records, while the user can also directly inspect and visit all \tool\ records in which the entity instance appears. 
The user is also able to filter the displayed entity instances based on a template type or record, as well as export them to Excel for further offline analysis.

The curated information in the vocabularies and the entity instances is exploited for offering better quantitative analysis and exploration services. Specifically, this data together with all data in the records are automatically transformed to RDF using the X3ML framework (cf.~Section~\ref{subsec:apprDataTransformation}). 
In X3ML, a \textit{schema mapping} file has to be created for i)~each different template, ii)~each type of entities, and iii)~the vocabularies. As we described in the previous section, a schema mapping indicates how the data must be mapped to elements of the target (domain) ontology. 
Finally, the transformed (and integrated) data are ingested in an RDF triplestore, which is the database of the ResearchSpace data exploration platform (cf.~Section~\ref{subsec:dataAnalysisExpl}).

\subsection{Technical Details}
\label{subsec:techDetails}
\tool\ is a web-based software system that runs on a web browser, either online or offline as a standalone application. 
It has an embedded web server, but may as well run using a static HTML page.
The system has been developed using Bootstrap, JavaScript, and jQuery. For managing the data entry forms, we make use of the JavaScript libary \textit{Handsontable}\footnote{\url{https://handsontable.com/}}.
All the data are saved in JSON format and stored locally on the web browser using multiple PouchDB\footnote{\url{https://pouchdb.com/}} databases. 
The local data are synchronised from time to time with the online \sq{central} data stored in multiple databases on an Apache CouchDB\footnote{\url{https://couchdb.apache.org/}} database management system.

In more detail, PouchDB is an open-source JavaScript database, inspired by Apache CouchDB, that is designed to run well within the browser. Its main advantage is easy replication from and to CouchDB and compatible servers. Every time user starts FastCat, the templates are replicated from the server if there is an internet connection. Otherwise, templates are retrieved from user’s local PouchDB database. Whenever user saves a record, it is saved locally in user’s local PouchDB database. If user is online when saving, record is also replicated to the server for safekeeping. In addition, if user has also decided to share a record publicly, then on save, information about terms and instances used inside the record are also replicated to the server to their corresponding databases.

\section{Use Case: Using \tool\ in SeaLiT}
\label{sec:usecase}

\tool\ is currently used in the SeaLiT project by around 30 users in 5 countries. 
The users are historians, like postdoctoral researchers, professors, Ph.D. students, and research assistants, of the following 5 organisations: 
i)~Institute for Mediterranean Studies, Foundation for Research and Technology - Hellas (Rethymno, Greece),  
ii)~T.I.G., University of Barcelona (Barcelona, Spain),
iii)~NAVLAB, University of Genoa (Genoa, Italy), 
iv)~UMR TELEMME, University Aix-Marseille (Aix-en-Provence, France), 
v)~Department of History, University of Zadar (Zadar, Croatia).

The number of configured templates is currently 20, representing 20 different types of data sources, and the total number of records is 605. Table \ref{tab:templates} gives the names of all templates and the corresponding number of records for each template (as of April 2021).
The transcribed data are written in 5 languages: Greek, French, Spanish (Castilian), Italian, and Russian. 
The template with the larger number of  records is \textit{Crew List (Ruoli di Equipaggio)} (98 records). This template contains the following 6 tables: 
\begin{itemize}
    \item \textbf{FastCat Record Information:} Table existing in all templates, containing metadata information about the record, in particular: id, creation date, last modification date, author name, author role. 
    \item\textbf{Source Identity:} Table existing in all templates, containing information about the data source, in particular: name and location of archive/library, register number, document number, date of release, name and location of issuing authority, source type, fond number and title, series number, file number and title.
    \item \textbf{Ship Identity:} Table containing information about the ship, in particular: ship name, ship type, tonnage, construction location, construction year, ammunition information (type, value, unit), registry port, registry number, owner. 
    \item \textbf{Crew List:} Table containing information about the crew members, in particular: embarkation port and date, discharge port and date, person (lastname, firstname, citizedship, location of residence, date of birth, serial number), profession/rank, wage (form of payment, value), total duration (months, days), pension fund (in Italian Lira). 
    \item \textbf{Documented Navigation:} Table containing information about the ship navigation, in particular: start and end date of navigation, total navigation duration (months, days), first planned destinations, total crew number (captain included).
    \item \textbf{Route:} Table containing information about the navigation route, in particular: departure port, departure date, note about the departure, arrival port, arrival date, note about the arrival. 
\end{itemize}
An example of a record belonging to this template is  available at \url{https://tinyurl.com/yxfbu5e5}.

\begin{table}[h]
    \centering
    \caption{List of \tool\ templates used in SeaLiT and corresponding number of records (as of April 2021).}
    \begin{tabular}{p{3mm}|p{100mm}|p{15mm}} 
        \toprule
        No & Template Name & \# Records \\
        \midrule
         1  &   Accounts book & 14\\
         2  &   Census La Ciotat & 63 \\
         3  &   Civil Register & 29 \\
         4  &   Crew and displacement list (Roll) & 35\\
         5  &   Crew List (Ruoli di Equipaggio) & 98 \\
         6  &   Employment records, Shipyards of Messageries Maritimes, La Ciotat & 50 \\
         7  &   First national all-Russian census of the Russian Empire & 6 \\
         8  &   General Spanish Crew List & 64 \\
         9  &   Maritime Register of the State for La Ciotat & 1 \\
         10 &   List of ships & 71 \\
         11 &   Logbook & 17 \\
         12 &   Naval Ship Register List & 2 \\
         13 &   Notarial Deeds & 10 \\ 
         14 &   Payroll & 7 \\ 
         15 &   Payroll of Russian Steam Navigation and Trading Company & 14 \\ 
         16 &   Register of Maritime personel & 4 \\ 
         17 &   Register of Maritime workers (Matricole della gente di mare) & 6 \\ 
         18 &   Sailors register (Libro de registro de marineros) & 52\\ 
         19 &   Seagoing Personel & 52 \\ 
         20 &   Students Register & 10 \\ 
        \bottomrule
    \end{tabular}
    \label{tab:templates}
\end{table}

In \toolteam, the total number of vocabularies is 52. Examples of vocabularies include: \textit{ship type, flag, marital status, religion, military service status, nationality, profession, reason of death, wind direction, wind strength}.
With respect to the entity instances, there are currently 76,643 person instances, 8,814 location instances, 2,354 ship instances and 1,197 legal entity instances. Note here that the manual instance matching process made by the historians (cf. Section \ref{subsec:datacuration}) is still undergoing, which means that the number of distinct entities (i.e., distinct entity identities) in each of the four entity categories is expected to be much lower at the end of the curation process (if not additional source are digitised).

\begin{figure}[h]
	\centering
	\includegraphics[width=\linewidth]{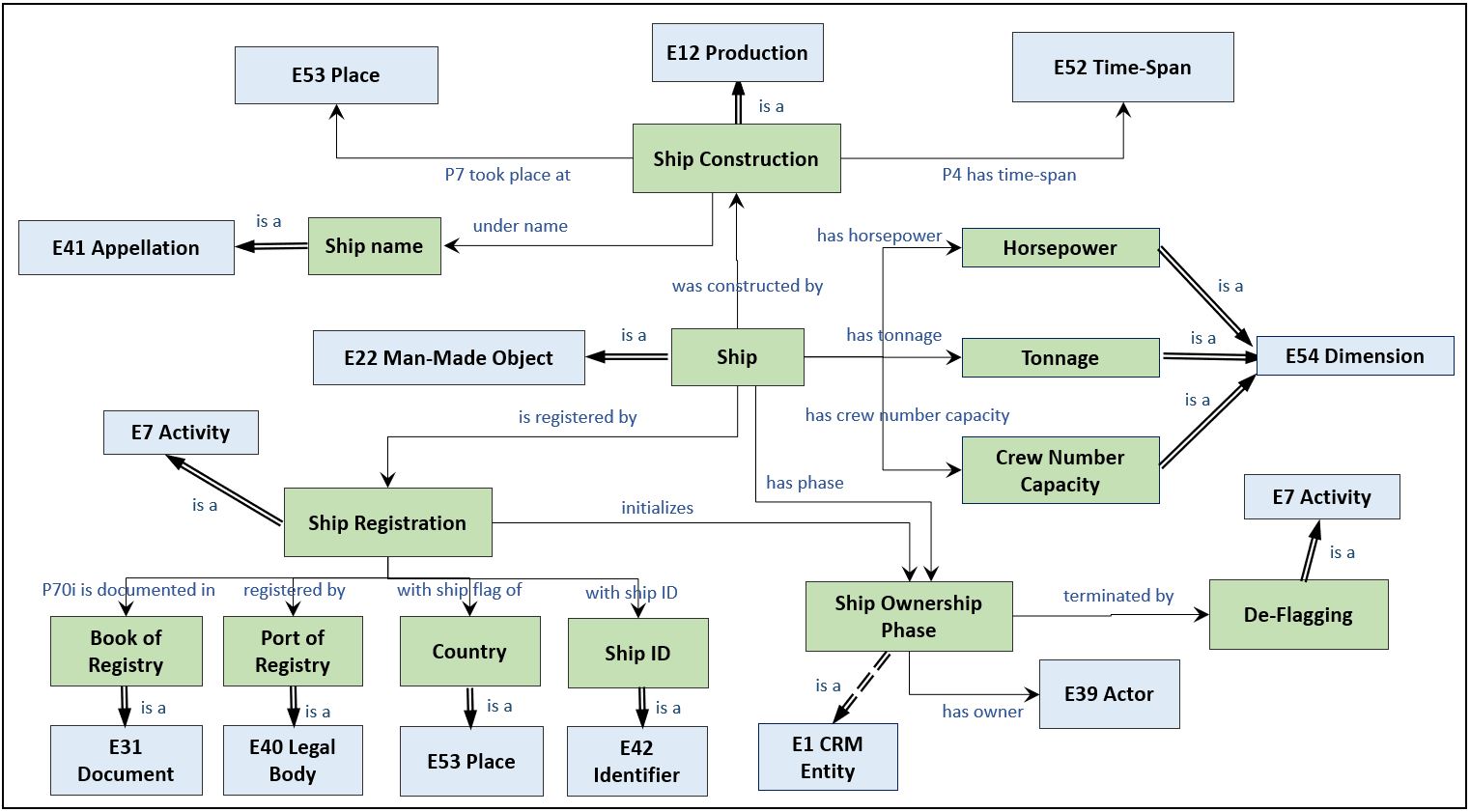}
	\caption{Modelling information about a \textit{ship}.}
	\label{fig:model}
\end{figure}

\begin{figure}[h]
	\centering
	\includegraphics[width=\linewidth]{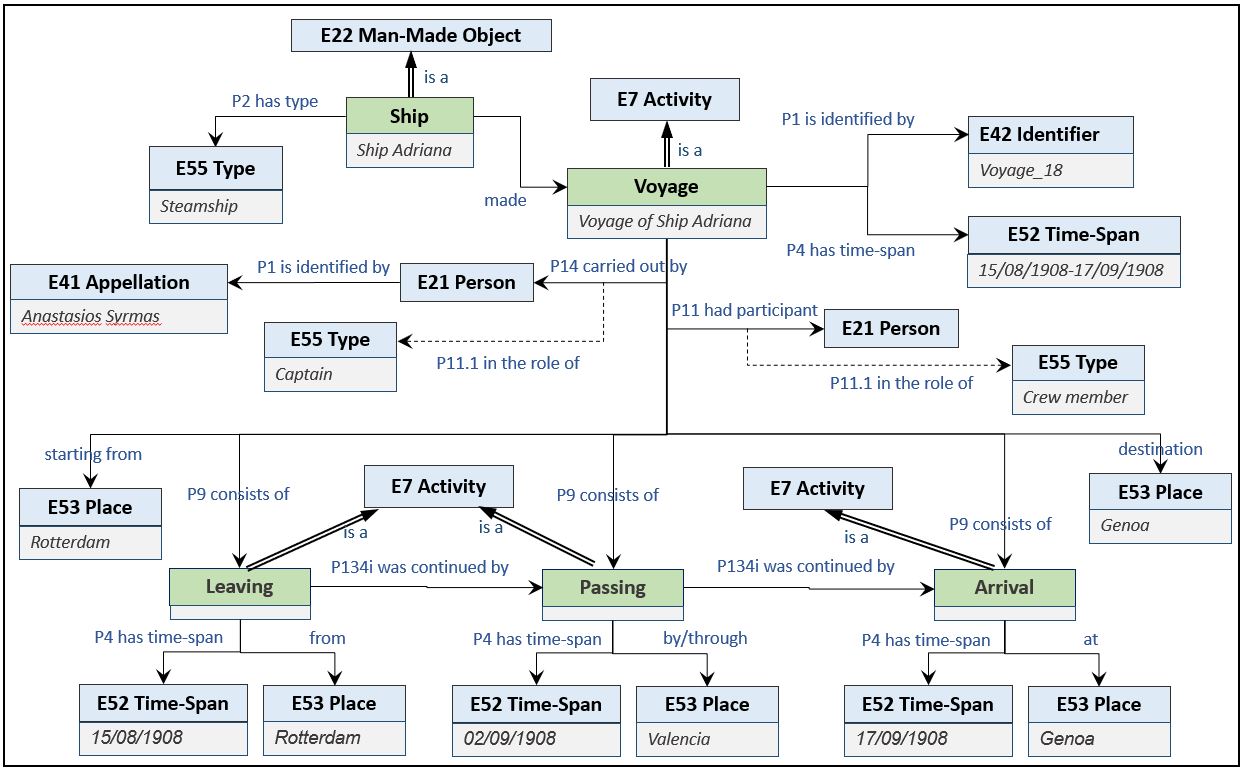}
	\caption{Modelling information about a \textit{ship voyage}.}
	\label{fig:model2}
\end{figure}

For transforming the curated data of the \tool\ records into a semantic network of linked data, we first created a data model compatible with CIDOC-CRM, which we call \sq{SeaLiT Ontology}. 
Figures~\ref{fig:model} and~\ref{fig:model2} illustrate a (small) part of the model, showing how information about a \textit{ship} and a \textit{ship voyage} is modelled. Each new class introduced in this model is subclass of a CIDOC-CRM class and, likewise, each property is sub-property of a CIDOC-CRM property. The data model is under constant evaluation as long as new \tool\ templates are created and analysed. 

\begin{figure}[h]
	\centering
	\includegraphics[width=\linewidth]{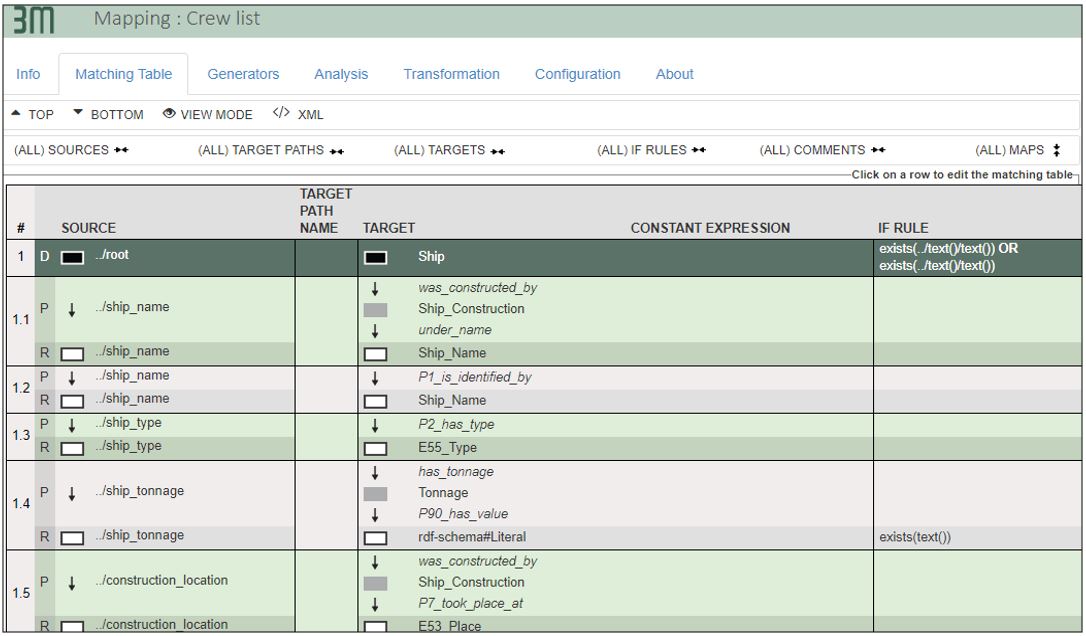}
	\caption{A part of a mapping definition in 3M Editor for the template \textit{Crew List (Ruoli di Equipaggio)}.}
	\label{fig:mappingsExample}
\end{figure}

\begin{figure}[h]
	\centering
	\includegraphics[width=\linewidth]{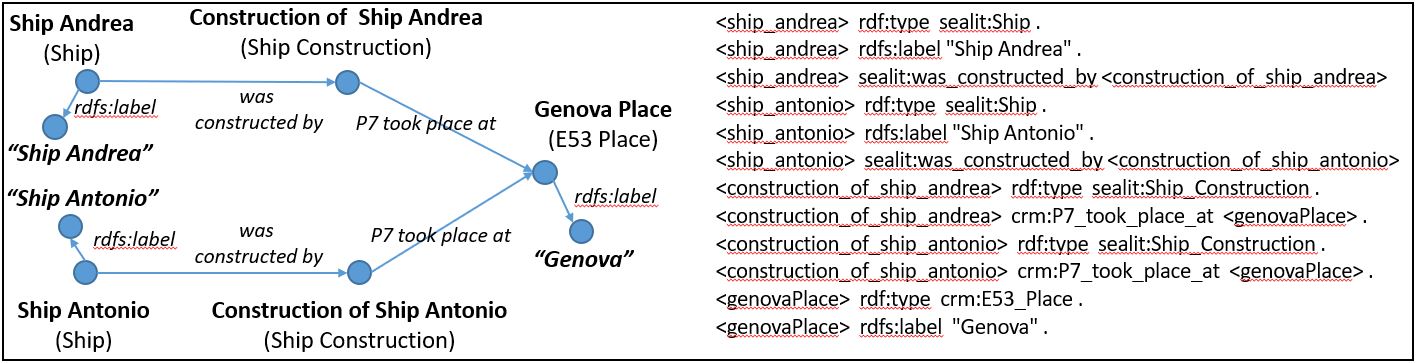}
	\caption{An example of (a part of) a semantic network and the corresponding RDF triples.}
	\label{fig:networkExample}
\end{figure}

Based on this data model, we create the schema mappings using X3ML and 3M Editor (as described in Section~\ref{subsec:apprDataTransformation}). In particular, we have to create: 
a) one mapping definition file for each distinct template in \tool\ (for transforming the record data), 
b) one mapping definition file for each category of entities in \toolteam\ (for transforming the curated entity data), and 
c) one mapping definition file for the vocabularies in \toolteam\ (for transforming the curated vocabulary terms). 
This sums to a total of (currently) 25 mapping definition files (20 for the templates, 4 for the four categories of entities, and 1 for the vocabularies). 
Also, a \textit{URI generation policy} file has been created which is required by X3ML for assigning URIs to the instances during the data transformation process.
Fig.~\ref{fig:mappingsExample} shows an example of a mapping definition file in 3M Editor which corresponds to the FAST CAT template \textit{Crew List (Ruoli di Equipaggio)}. 
We see, for example, that the field \textit{construction location} of the source schema (mapping \#1.5 in the figure), corresponding to a particular table column in the template, is mapped to the following path in the data model (cf. Fig.~\ref{fig:model}): \textit{Ship---was constructed by---Ship Construction---P7 took place at---E53 Place}. 
Considering this mapping definition and also that a \tool\ record contains, for instance, the information that the ships \textit{Andrea} and \textit{Antonio} have both construction location \textit{Genova}, the derived semantic network and the corresponding RDF triples are those displayed in Fig.~\ref{fig:networkExample}. 

\section{Conclusion}
\label{sec:conclusion}

We have focused on the common data management challenges faced by many researchers in descriptive and empirical sciences like History.
Our particular focus is on research that requires some type of quantitative analysis in which the data can be modelled in tabular form and comes from multiple sources/providers, which means that data consolidation is required in order to enable effective analysis and exploration services that combine information of different sources.
To this end, we described a general data management methodology that focuses on semantic interoperability, long-term validity, and making the data (re)usable beyond the context of a particular research problem or project. 
This motivated the development of \tool,
a collaborative system for data entry and curation in Digital Humanities and beyond.
We have presented the functionality and interface of \tool\
that supports \textit{data transcription} (with emphasis on data entry speed) and  \textit{data consolidation} (correction, normalisation, multi-level instance matching, ontology-based data transformation).
\tool\ is innovative in its ability to support features like nested tabular structures for data entry, embedded instance matching and vocabulary maintenance processes, and provenance-aware data curation.

The \tool\ system as well as the data management approach described in this paper are currently in use by historians working in the context of \textit{SeaLiT}, a European research project (ERC) of maritime history, for the transcription and curation of historical archival sources like handwritten crew lists, civil registers and logbooks.
Nevertheless, the overall data management methodology is configurable which means that it can be used for transcribing, curating, transforming and then exploring other data sources, beyond the case of maritime history. 
The actual steps for applying this methodology on another domain are the following:
i)~creation of the \tool\ templates (i.e., decide on the structure of the tables, the value type of each table column, the entity categories and the vocabularies);
ii)~selection and/or design of the domain ontology to use for semantically representing the data of the \tool\ records; 
iii)~creation of the schema mappings and the URI generation policies for each template and each category of entities (which will be used together with the domain ontology for producing the semantic network / RDF data).  

We are currently working on how to make easier (and more standardised) the configuration of \tool, in particular on how to enable the straightforward creation or change of a \tool\ template. 
Our long-term objective is to enable plain users (like historians) to create their own templates, or modify existing ones, through a user-friendly environment within \tool. 
Another interesting direction for future work is the integration of the domain ontology used for data transformation within the structure of the templates, 
e.g., by associating each table column to a concept or relationship in the ontology. 
That would  facilitate 
the definition of the mapping in X3ML (which is a quite time consuming task) 
and  enable better handling of the
updates in the schema when there are changes in the templates.

\section*{Acknowledgements}

This work has received funding from the European Union's Horizon 2020 research and innovation programme under 
i)~the Marie Sklodowska-Curie grant agreement No 890861 (Individual Fellowship, Project \textit{\q{ReKnow - Research Knowledge Documentation, Analysis and Exploration in Empirical and Descriptive Sciences}}), 
and ii)~the European Research Council (ERC) grant agreement No 714437 (Project \textit{\q{SeaLiT - Seafaring Lives in Transition. Mediterranean Maritime Labour and Shipping during Globalization, 1850s-1920s}}).